# Instrument for *in situ* hard x-ray nanobeam characterization during epitaxial crystallization and materials transformations

Samuel D. Marks,[1,†] Peiyu Quan,[1,†] Rui Liu,[1] Matthew J. Highland,[3] Hua Zhou,[3] Thomas F. Keuch,[1,2] G. Brian Stephenson,[4] and Paul G. Evans[1,*]

[1] *Department of Materials Science and Engineering, University of Wisconsin, Madison, WI 53706 USA*

[2] *Department of Chemical and Biological Engineering, University of Wisconsin, Madison, WI 53706 USA*

[3] *X-ray Science Division, Argonne National Laboratory, Argonne, Illinois 60439 USA*

[4] *Materials Science Division, Argonne National Laboratory, Argonne, Illinois 60439 USA*

**ABSTRACT**

Solid-phase epitaxy (SPE) and other three-dimensional epitaxial crystallization processes pose challenging structural and chemical characterization problems. The concentration of defects, the spatial distribution of elastic strain, and the chemical state of ions each vary with nanoscale characteristic length scales and depend sensitively on the gas environment and elastic boundary conditions during growth. The lateral or three-dimensional propagation of crystalline interfaces in SPE has nanoscale or submicron characteristic distances during typical crystallization times. An *in situ* synchrotron hard x-ray instrument allows these features to be studied during deposition and crystallization using diffraction, resonant scattering, nanobeam and coherent diffraction imaging, and reflectivity. The instrument incorporates a compact deposition system allowing the use of short-working-distance x-ray focusing optics. Layers are deposited using radio-frequency magnetron sputtering and evaporation sources. The deposition system provides control of the gas atmosphere and sample temperature. The sample is positioned using a stable mechanical design to minimize vibration and drift and employs precise translation stages to enable nanobeam experiments. Results of *in situ* x-ray characterization of the amorphous thin film deposition process for a $SrTiO_3$/$BaTiO_3$ multilayer illustrate implementation of this instrument.

## I. INTRODUCTION

Epitaxial crystallization involves kinetic processes that occur with rates that depend strongly on temperature, stress, gas environment, and other variables. In materials systems ranging from semiconductors to complex oxides, these processes result in the development of structural features with nanoscale to sub-μm dimensions, including dislocations associated with plastic relaxation and the formation of quantum dots and other surface nanostructures.[1-3] The ionic valence and phase in epitaxial metal oxides similarly depend on the composition of the gas environment during growth and crystallization.[4] The development of microstructural and chemical features can be probed with focused hard x-ray beams using scanning diffraction microscopy, ptychography, and coherent diffraction imaging.[5-7] Focused x-ray probes can be used to form images of the distribution of dislocations, the formation of competing polymorphs, phase separation, and local elastic distortions, as seen, for example, in relaxed thin-film layers.[8] *In situ* x-ray scattering provides insight with time resolution that matches the characteristic timescales of the relevant processes, but has so far been largely limited to studies of structural parameters over large areas beyond micro-scale.[9, 10] Instruments that combine focused hard x-ray probes with thin film growth environments have the potential to provide significant insight into epitaxial crystallization processes.

We focus here on the development of instrumentation that addresses challenges associated with the crystallization of complex oxides via SPE and related processes yielding three-dimensional nanoscale crystalline geometries. *In situ* instruments employing synchrotron x-ray radiation to study epitaxial crystallization have been largely designed to characterize thin film epitaxy using x-ray diffraction, grazing-incidence small angle scattering, x-ray reflectivity, and coherent x-ray scattering.[11-14] For example, the application of *in situ* crystal growth instruments to the epitaxial growth of oxides has allowed x-ray reflectivity, grazing-incidence diffuse x-ray scattering, and crystal truncation rod analysis to be used to investigate the growth morphology of planar complex oxide thin films.[15-20] These instruments, however, do not accommodate short-working-distance x-ray optics for nanobeam diffraction or imaging experiments and thus provide structural information sampling micron-scale or larger regions of planar thin films.

Here we describe instrumentation enabling the investigation of SPE and other epitaxial synthesis processes using focused synchrotron x-ray probes for *in situ* characterization. The deposition system incorporates a vacuum chamber designed specifically to facilitate the use of x-ray focusing optics and coherent x-ray techniques. The layers can be deposited by radiofrequency (RF) sputtering and vacuum evaporation. The crystallization occurs via post-deposition sample heating at controlled temperatures and in an environment of controlled gas composition and pressure. The instrument described here is integrated into the specialized diffractometer system for *in situ* coherent x-ray studies of materials dynamics[20] located at station 12-ID-D of the Advanced Photon Source (APS) at Argonne National Laboratory.

Combining nanoscale x-ray beams with *in situ* deposition and crystallization capabilities imposes significant constraints in the design of the instrument. Hard x-ray Fresnel zone plate focusing optics, for example, require short working distances on the order of 10 cm for the zone plate and of a few cm for the order sorting aperture that selects the primary focused beam.[21, 22] The deposition chamber described here allows x-ray optical elements mounted in air with short, few-cm, working distances. The design imposes the minimum possible restriction on the angular range available for x-ray diffraction and scattering studies. The vacuum windows incorporated in the instrument are mounted on recessed conical flanges to allow for a wide range of incident- and scattered-beam directions, making a large volume of reciprocal space accessible in experiments. The low level of sample vibration required to employ sub-micron beams is achieved with mechanically rigid sample stage and input optics platform. The mechanical design and geometry of the deposition system are described below, including studies testing the deposition capabilities and the mechanical stability during rotation with *in situ* reflectivity and diffraction measurements. The characterization of nanobeam methods employing this instrument will be reported in a subsequent manuscript.

## II. INSTRUMENT DESIGN

The instrument described here has been designed for studies using several synchrotron x-ray techniques: (i) x-ray diffraction and scattering, including nanobeam diffraction microscopy, coherent diffraction imaging, truncation rod analysis, and diffuse scattering; (ii) x-ray reflectivity; and (iii) x-ray photon correlation spectroscopy. The instrument is configured for station 12-ID-D of the APS, which employs an undulator source to provide synchrotron radiation at photon energies from 4.5 to 40 keV. The photon flux in an unfocused $1 \times 1$ mm$^2$ beam at station 12-ID-D is $10^{13}$ photons s$^{-1}$ at 10 keV. A linear compound refractive lens can focus the beam vertically to 4 μm for experiments probing large areas.[23] The energy resolution provided by the Si (111) monochromator is $\Delta E/E = 1.31 \times 10^{-4}$.

The overall design of the instrument consists of two modules: a deposition system mounted on a diffractometer and separate set of input x-ray optics in air on a breadboard, as shown in Fig. 1(a). The systems are incorporated as modules of the versatile hexapod-based diffractometer at the APS station 12-ID-D.[20] The modular design enables rapid changeover between experiments by minimizing the reconfiguration of the beamline before and after experiments.

**A. Mechanical design and configuration**

The deposition system is shown schematically in Fig. 1(b). The x-ray entry and exit vacuum windows are Be foils with thicknesses of 0.18 mm and diameters of 4.57 cm, providing transmission of 83% per window at normal incidence at the 4.5 keV minimum energy of station 12-ID-D. Above 15 keV, the transmission for each window at normal incidence is greater than 99%. The x-ray windows are mounted on DN300 flanges with cones that are recessed into the chamber, with the faces of the windows at distances of 2.4 cm from the sample, as shown in Fig. 1(c). The conical opening angle of the Be windows is 98°, making a large volume of reciprocal space accessible for diffraction studies.

The deposition system is connected to the final axis of rotation of the diffractometer (phi) via a differentially pumped rotary seal (RNN-400, Thermionics Laboratory, Inc.), component 7 in Fig. 1(b), which enables free phi rotation of the sample with respect to the rest of the deposition system. The top flange of the rotary seal is incorporated in the deposition system. A locking mechanism is then used to secure the stationary end of the rotary seal to the hexapod that provides the other sample tilts and translations so that the deposition system does not rotate with the sample phi motion. The electronics and the 700 W thermoelectric water chiller for the deposition system (Thermotek 3U, Cole-Parmer Instrument Company, LLC) are held on a cart that is situated adjacent to the diffractometer during experiments.

High-precision diffraction measurements require placing and maintaining the sample accurately near the center of rotation to reduce systematic angular errors and to keep the beam footprint at the same location within the sample during angular scans. A 19 mm outer-diameter stainless-steel cylinder with a wall thickness of 2 mm connects the sample to a 3-axis translation stage mounted outside the vacuum and connected to the chamber by a thin-wall bellows. The cylinder supporting the sample is with its long axis along a vertical direction in order to avoid coupling between vertical vibrational displacements to cantilever vibrational modes. The three-axis translation stage rotates with the final axis of rotation, ensuring that the sample region of interest can be positioned at the center of rotation of the diffractometer. The angular reproducibility of this stage is demonstrated below in reflectivity and diffraction experiments. Other angular degrees of freedom are provided by the hexapod on which the deposition system is mounted.[20]

**B. Deposition system vacuum and gas environment control**

The deposition system uses metal seals permitting ultra-high vacuum (UHV) pressures after baking. The deposition system is evacuated using a turbomolecular pump (TMP) (HiPace 300, Pfeiffer Vacuum, Inc.) mounted horizontally on a DN100 flange. The TMP is rigidly fixed to the deposition system. Multiple TMP rotation speed settings are available for different operating conditions. The base pressure of the deposition system with differential pumping of the rotary seal and without baking is $10^{-8}$ Torr. The pressure is measured with a cold cathode gauge. The deposition system is vented with dry $N_2$ gas using a manual vent valve.

**C. Materials deposition and sample geometry**

The deposition system includes two methods for thin-film deposition: (i) up to two RF magnetron sputtering sources that can be configured for either on- or off-axis deposition geometries and (ii) a single thermal evaporation deposition source. The RF magnetron sputtering sources (Model A320, AJA International, Inc.) employ 5.08 cm-diameter targets positioned with a working distance of 5 cm from the sample. The sputtering sources are cooled with water. RF power supplies (0613 GTC, T&C Power Conversion) operating at 13.56 MHz provide the power to generate a plasma at each source, with an output power that can be tuned from 1 to 200 W. The gas environment during deposition is controlled with an external gas manifold that integrates mass flow controllers (MKS, Inc.) and shutoff valves. The flow rates for process gases are controlled using a vacuum controller (Series 946, MKS) and set during growth with a throttling valve mounted to the TMP or by adjusting gas flow rates until the desired process pressure is reached. Each deposition source is equipped with an independently controlled electropneumatic shutter.

The deposition system can be configured for either vertical or horizontal surface normal geometries. In horizontal geometry experiments, the surface normal of the sample and the incident x-ray beam define a plane that is horizontal, as shown in Fig. 2(a). In the horizontal configuration, the sputter guns can be mounted in ports providing on-axis deposition using port 2A or 60° off-axis using port 2B. The horizontal configuration also permits the use of a thermal evaporator at 45° off-axis using port 3.

In the vertical geometry, shown in Fig. 2(b), the plane defined by the surface normal of the sample and the incident x-ray beam is vertical. This configuration positions the sample so that sputtering sources have an approximately 30° off-axis sputtering geometry, using ports 2C and 2B in Fig. 1(b), or a 90° off-axis geometry using port 2A.

**D. Sample heating and exchange**

Samples are mounted to the surface of a resistive heater encapsulated in a 12.7 mm diameter cylindrical ceramic block (Heatwave Labs, Inc.), shown in Fig. 3(a). Samples can be heated up to 1000 °C with ramp rates up to 10 °C/s. The samples are held to the surface through mechanical clips oriented to be clear of the incident and scattered x-ray beams. A boron nitride ceramic post, element 3 in Fig. 3(b), provides a thermal barrier between the heater mount and the sample heater. The temperature of the sample is measured with a type-K thermocouple adjacent to the sample. The temperature of the heater is independently monitored with a second type-K thermocouple in contact with the rear face of the heater.

The samples are exchanged by removing DN100 flange on the deposition chamber and extracting the sample stage. Vacuum-side thermocouple and electrical connections to the sample heater remain intact during sample exchange. The sample stage incorporates a kinematic mount that serves as the detachment point during sample exchange. Fig. 3(b) shows an image of the interior of the deposition system captured during a multilayer thin film deposition using the horizontal scattering geometry. The perspective in Fig. 3(b) shows the x-ray windows covered in protective Al foil, an active sputter deposition source with accompanying plasma and open shutter occupying port 2A, and an active sputter deposition sources with a closed shutter at port 2B. The sample and sources are viewed through a viewport at port 2C, with sample stage viewed from behind.

## III. APPLICATIONS IN COMPLEX OXIDE SOLID-PHASE EPITAXY: REFLECTIVITY AND DIFFRACTION

The capabilities of this instrument were demonstrated in a series of deposition and x-ray characterization experiments involving the growth of amorphous oxide layers, x-ray diffraction, and reflectivity. The experiments focus on the deposition and characterization processes relevant to the crystallization of complex oxides via SPE. SPE involves the creation of amorphous oxide layers and their subsequent epitaxial crystallization via heating, electron-beam irradiation, or mechanical distortion.[24] The kinetic processes of SPE involve separate steps for deposition and crystallization and thus can, in many cases, employ crystallization temperatures that minimize competing processes such as re-evaporation, interface reaction, and phase separation.[25, 26] Beyond planar epitaxial layers, SPE can be employed in the three-dimensional (3D) epitaxial growth of oxides in intricate pre-patterned structures or laterally from nanoscale seeds, providing a method to produce thin films and devices in new geometries.[27, 28] The region crystallized by SPE has an orientation that can be templated either by a single-crystalline substrate or nanocrystalline seeds.[29, 30]

The experiments reported here were designed to evaluate the positioning and alignment of the sample and *in situ* characterization of the layer structure developed during deposition. The incident beam had an x-ray photon energy of 12.13 keV. These experiments employed the horizontal surface normal geometry shown in Fig. 2(a). The intensity of scattered x-rays was recorded using a pixel array detector (LAMBDA 250K, X-Spectrum GmbH) with 55-µm-square pixels located 331 mm from the sample.

Layers of $SrTiO_3$ (STO) and $BaTiO_3$ (BTO) were deposited on a single crystal STO (001) substrate. The STO and BTO sputtering sources were mounted on-axis on port 2A and 60° off-axis on port 2B, respectively. The sources were simultaneously excited by RF power and the shutters in front of each source were controlled to select each source for deposition. As described below, the deposition rate of $SrTiO_3$ (STO) at 30 W and 40 mTorr Ar is 3.1 nm/hour in on-axis sputtering geometry. The BTO deposition rate under identical gas and power conditions was lower because of the 60° off-axis sputtering geometry, 0.5 nm/hour. The deposition system was initially evacuated to $5 \times 10^{-6}$ Torr before flowing 450 sccm Ar and 50 sccm $O_2$ to increase the pressure to 45 mTorr. A pre-sputtering procedure was employed to prepare the STO and BTO targets for deposition by initiating a plasma at 30 W with the shutters closed. The voltages for the STO and BTO guns during the pre-sputtering were 18 V and 36 V, respectively.

X-ray reflectivity and growth oscillations were measured to demonstrate the *in situ* characterization capabilities of the instrument. Diffraction rocking curves of the substrate Bragg reflection were measured to test the repeatability and stability of the sample rotation.

### A. *In situ* x-ray reflectivity

Measurements of the specularly reflected x-ray intensity as a function of wavevector were acquired at multiple stages during the deposition. The intensity distribution in the reflectivity profiles provides information about the layer thickness, surface and interface roughness, and the density of the amorphous layers. Reflectivity scans can be conducted with time resolution that depends on the angular range of the scan and the counting time. The reflectivity scans for this study were collected with 201 points in angular steps of 0.0125° and required less than 5 min per scan. This time resolution allows important structural aspects during the deposition process to be captured, including the formation of interfaces, the evolution of surface and interface roughness, and layer thickness.

X-ray reflectivity scans were collected at specific times during the deposition of the STO/BTO multilayer, at configurations depicted in Fig. 4(a). The conditions were (i) the STO substrate before deposition, (ii) after the deposition of BTO for 315 min and STO for 10 min, and (iii) after a further 187 min of STO deposition. The corresponding x-ray reflectivity curves are shown in Fig. 4(b), shown as a function of $Q_z$, the x-ray wavevector along the surface normal direction. The scans were acquired without pausing the deposition process, resulting in an uncertainty in the STO layer thickness in each case of 0.3 nm.

The x-ray reflectivity curves were interpreted using a model consisting of the top STO layer, a BTO layer, and the STO substrate. The density of crystalline STO substrate was fixed at 5.1 g/cm$^3$. The densities of amorphous STO and BTO were determined from the analysis of curve (ii), with values of 4.2 g/cm$^3$ and 4.5 g/cm$^3$, respectively. The thicknesses obtained from the fits were, for curve (ii) a BTO thickness of 2.4 nm and an STO thickness of 0.6 nm and, for curve (iii), an additional 9.5 nm of STO. The best fit in the simulations included a 0.3 nm root-mean-square (rms) roughness at the interface between the BTO and STO layers, and 0.4 nm rms at the surface of the STO layer. The deposition rates for STO and BTO determined from the deposition times and layer thicknesses determined from Fig. 4(b) were 3.1 nm/hour and 0.5 nm/hour, respectively.

A second approach to *in situ* characterization of amorphous layer deposition involved monitoring the time dependence of the reflected x-ray intensity at a single wavevector to provide a continuous measurement of the deposited thickness. The reflected intensity was measured during sequential depositions of topmost STO and BTO layers with durations of 20 min each. These layers were deposited on a preexisting STO/BTO multilayer, as shown in Fig. 5(a).

The time dependence of the reflected x-ray intensity at $Q_z = 0.499$ Å$^{-1}$ is shown in Fig. 5(b). The wavevector selected for this measurement was greater than the critical angle for total external reflection and small enough to capture clear oscillations in the reflected x-ray intensity. The time interval of Fig. 5(b) spans the deposition of a layer of STO followed by a BTO layer and then by a further STO layer. The reflected intensity in Fig. 5(b) is normalized to 1 at its local maximum value during this time. Three segments of distinctly varying reflected intensity are apparent during the deposition of the STO and BTO layers. Segments (i) and (iii) occur during the deposition of STO and exhibit oscillations with a period of 23 min. Segment (ii) corresponds to deposition of BTO. The time variation of the intensity in segment (ii) is much slower than in segments (i) and (iii) because the deposition rate of BTO is significantly lower than for STO.

Insight into the time dependence of x-ray reflectivity during deposition can be gained from a model of the reflectivity of a single layer of continuously increasing thickness. For a layer of thickness $d$, the amplitude reflectivity $r$ at wavevector $Q_z$ is:[31]

$$r = -\frac{4\pi\rho r_0}{Q_z^2}(e^{iQ_z d} - 1). \tag{1}$$

Here $r_0$ is the classical radius of the electron and $\rho$ is the electron density of the layer. The intensity reflectivity is the square magnitude of the amplitude reflectivity:

$$R = |r|^2 = 2\left(\frac{4\pi\rho r_0}{Q_z^2}\right)^2 (1 - cosQ_z d). \tag{2}$$

The reflectivity given by equation (2) is periodic such that successive maxima occur after increases in thickness $\Delta d$ satisfying $Q_z \Delta d = 2\pi$. With a constant deposition rate $v$ the increase in layer thickness after time $\Delta t$ is $\Delta d = v\,\Delta t$, where $v$ has units of thickness per unit time. The intensity oscillation period $\tau$ is:

$$\tau = \frac{2\pi}{Q_z v}. \tag{3}$$

The observed period of the oscillations in segments (i) and (iii) of Fig. 5(b) is 23 min, corresponding to a deposition rate of 3.1 nm/hour, matching the results from the full $Q_z$-dependent reflectivity curves shown in Fig. 4.

The results obtained from considering a single layer are also consistent with simulation considering the entire set of STO and BTO layers. The intensity as a function time at fixed $Q_z$ was simulated using the Parratt formalism including the thicknesses all of the layers and the roughness of their interfaces.[31, 32] The roughness at each interface and scattering length densities determined with *in situ* reflectivity measurements were used in the simulation of the time-dependent reflected intensity. Deposition rates for STO and BTO of 3.1 and 0.5 nm/hour, respectively, were used in the simulation. Figure 5(c) shows the simulated reflected intensity at $Q_z$ = 0.499 Å$^{-1}$ during the growth an STO/BTO/STO multilayer matching the experiment. The intensity oscillation period in Fig. 5(c) is 23 min during segments (i) and (ii), matching the experimental result.

The intensity oscillations observed in x-ray reflectivity due to increasing thickness are distinct from the growth oscillations observed in the specular intensity in reflection high-energy electron diffraction (RHEED). RHEED probes a nm-scale thickness near the surface exhibits oscillations during layer-by-layer growth due to the repeated monolayer-scale structural reconfiguration of the surface, with period equal to the size of the unit cell.[33] In comparison, the x-ray reflectivity intensity oscillations arise from the change in thickness and have a thickness and time dependence on $Q_z$. For example, at the midpoint between the Bragg reflections, often termed the anti-Bragg condition, where $Q_z=\pi/a$ for lattice parameter *a*, the x-ray intensity oscillations have period $\Delta d=2a$, different from the RHEED result. The *in situ* information provided by continuous monitoring of the x-ray reflectivity thus provides complimentary information, such as the thickness of amorphous layers.

## B. X-ray diffraction

The reproducibility of the rotation of the sample through the differentially pumped rotary seal was characterized by acquiring repeated rocking curve scans of the STO 001 substrate x-ray reflection. Figure 6(a) shows the diffracted intensity as a function of incident angle for three rocking scans around the STO 001 Bragg reflection at room temperature on which no thin film layers had been deposited. The full width at half maximum of the rocking curves is 0.015°. The maximum fractional difference in intensity at each angle among the three rocking scans is shown in Fig. 6(b). The difference in the intensities upon repeated scans illustrates no systematic peak shift and has maximum fractional difference of 0.04. The angular shift between scans is less than 0.001°. The negligible angular and peak intensity variation among the three rocking scans in Fig. 6(a) demonstrates accurate and repeatable sample rotation, indicating low radial runout and wobble in the rotation stage. This stability is critical for future experiments employing *in situ* nanobeam and coherent diffraction measurements where small deviations in sample position or alignment could corrupt the expected signal.

## IV. CONCLUSION

The x-ray diffraction and scattering instrument described here allows studies of *in situ* processes during deposition and crystallization. The instrument is optimized for the mechanical stability and short working distances required for *in situ* synchrotron x-ray nanodiffraction studies of materials transformations during synthesis. Wide-angle Be vacuum windows that are recessed toward the sample stage provide space for short-working distance x-ray focusing optics that can produce sub-µm size x-ray probes while facilitating a wide range of scattering geometries. Surface layers are deposited with multiple vapor phase deposition capabilities, including on- and off-axis RF magnetron sputtering for oxide materials, and a vacuum evaporation cell for high melting point elements. Key aspects of the deposition system with *in situ* characterization of the deposition of thin amorphous layers and crystallographic measurements of a single crystal STO substrate.

With the ongoing upgrade of the APS to incorporate a low-emittance multibend-acromat storage ring to produce synchrotron x-ray beams with orders-of-magnitude higher brilliance, we anticipate that there will be new capabilities in x-ray diffraction imaging with higher numerical aperture x-ray optics. Significant enhancement of the coherence of the upgraded APS will further enable coherent diffraction imaging and scattering studies of nano-scale dynamics that characterize the formation of thin epitaxial layers. The capabilities of this instrument, combined with the enhancements of the synchrotron x-ray probes at the APS, are particularly well suited for investigations of vertical and lateral solid phase epitaxial growth processes which are characterized by the propagation of an amorphous-crystalline interface, and for which there is relatively little experimental insight into key nanoscale phenomena associated with the solid-solid transformation.

[†] Equally contributing authors.

* Author to which correspondence should be addressed: pgevans@wisc.edu

**ACKNOWLEDGMENTS**

This research work was primarily funded by the NSF Division of Materials Research through the University of Wisconsin Materials Research Science and Engineering Center (Grant DMR-1720415). This research used resources of the Advanced Photon Source, a U.S. Department of Energy (DOE) Office of Science User Facility operated for the DOE Office of Science by Argonne National Laboratory under Contract No. DE-AC02-06CH11357. Integration of the instrument into the specialized diffractometer system for *in situ* coherent x-ray studies of materials dynamics was supported by the DOE Office of Science, Basic Energy Sciences, Materials Science and Engineering division.

**Data Availability**

The data that support the findings of this work are available from the corresponding author upon reasonable request.

**Figure 1** (a) *In situ* epitaxial crystallization instrument with x-ray focusing optics and deposition system. (b) Components of the deposition system: (1) recessed Be windows, (2) ports for mounting RF magnetron sputter deposition sources, with sources shown on 2A and 2B and a viewport for alignment and sample exchange on 2C, (3) port for thermal evaporation sources, (4) TMP, (5) thermocouple, gas, and power feedthroughs, (6) vacuum gauge, (7) differentially pumped rotary seal, (8) sample post, (9) mounting interface to diffractometer, and (10) sample translation stages. (c) Arrangement of the sample and of x-ray windows with the 2.4 cm separation between the center of the sample and the interior face of the window. (d) Deposition system installed at station 12-ID-D of the APS. (e) Perspective of the deposition instrument shown in (d).

**Figure 2** Configuration of the deposition instrument for (a) horizontal and (b) vertical sample normal geometries. The directions of the sample surface normal and the directions of the incident and specular x-ray beams are shown for each geometry. The incident directions of material from sputter deposition sources mounted on ports 2A, 2B, and 2C, labeled A, B, and C, respectively, are indicated.

**Figure 3** (a) Sample stage and heater. Mechanical clips (1) secure the sample to the face of the ceramic resistive heater (2), attached to a boron nitride post (3). (b) Interior of the deposition system during sputter deposition viewed through a vacuum window at port 2C.

**Figure 4** (a) Three stages of deposition for a STO/BTO multilayer film at which x-ray reflectivity scans were performed, including (i) prior to deposition, (ii) after 2.4 nm of BTO and 0.6 nm of STO, and (iii) after an additional 9.5 nm of STO. (b) X-ray reflectivity curves offset in intensity by a factor of $10^{1.5}$ each. The solid lines are the reflectivity simulated with the parameters given in the text.

**Figure 5** (a) Layer sequence of the amorphous STO/BTO multilayer thin film. (b) X-ray reflectivity measured at $Q_z = 0.499$ Å$^{-1}$ during the deposition of the topmost layers in (i)-(iii), normalized to the time point with the highest reflectivity. Colored bars below the curve represents periods of STO and BTO deposition, respectively. (c) Simulated reflectivity at $Q_z = 0.499$ Å$^{-1}$ as a function of deposition time plotted (black curve, left axis) during the STO/BTO multilayer deposition, normalized as in (b). The total deposited film thickness is also plotted (blue curve, right axis).

**Figure 6** (a) Rocking curve scans of the 001 Bragg reflection of the STO substrate. The integrated diffracted intensity collected from a three repeated rocking scans at room temperature prior to deposition, is plotted as a function of deviation from the Bragg angle, $\Delta\theta_B$. An offset of 2 x $10^5$ is applied to scans 2 and 3 for clarity. (b) The maximum fractional difference in diffracted intensity among the rocking scans in (a) at each angular step.

**Marks *et al.* Figure 1**

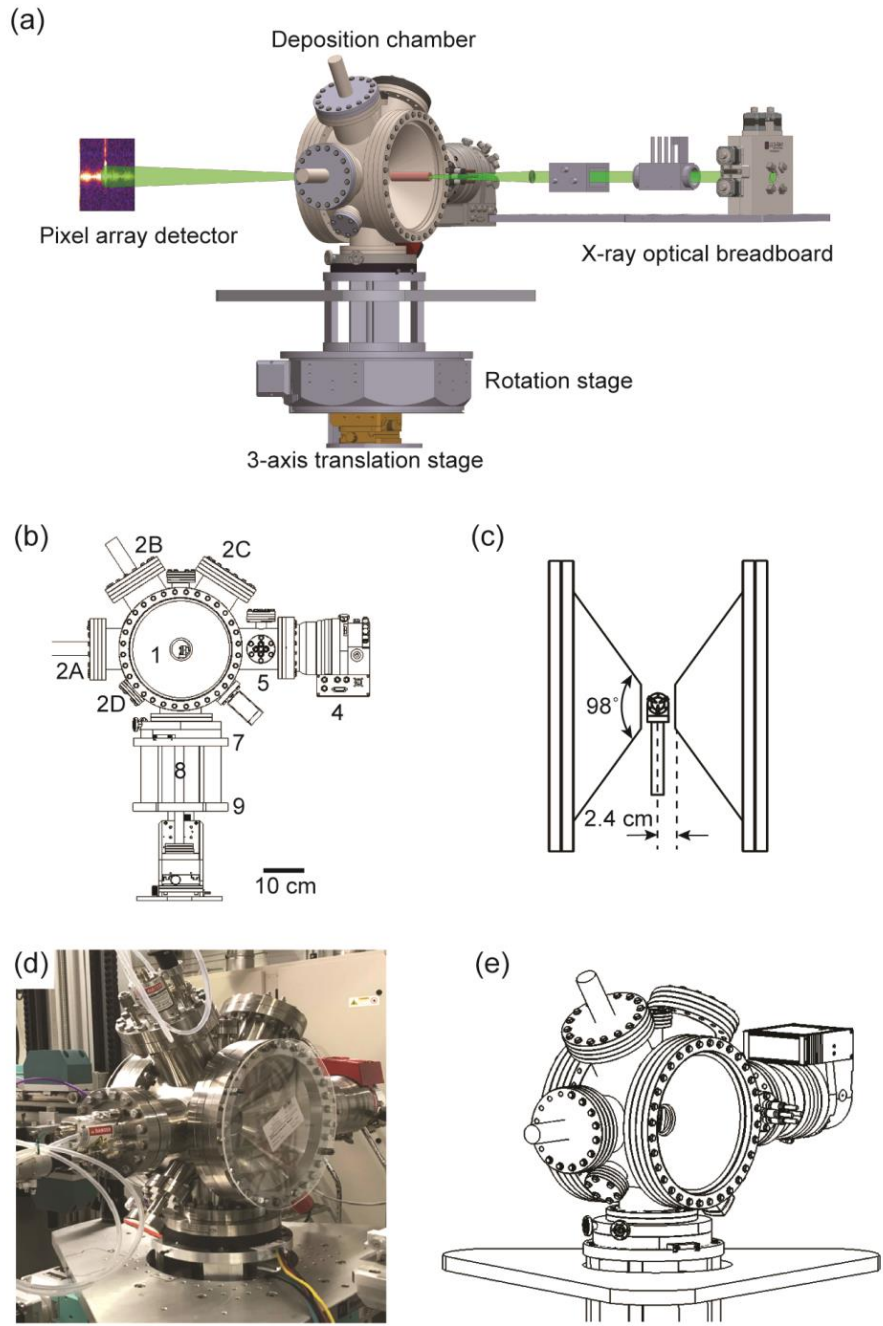

**Marks *et al.* Figure 2**

(a) Horizontal geometry

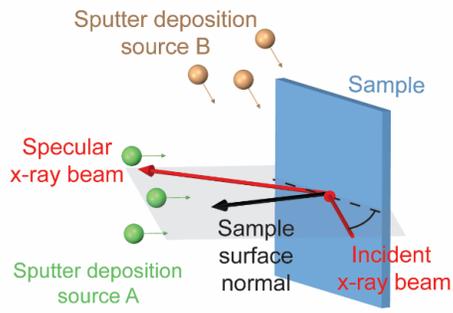

(b) Vertical geometry

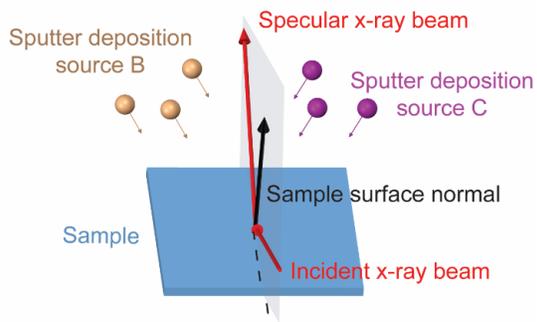

**Marks *et al.* Figure 3**

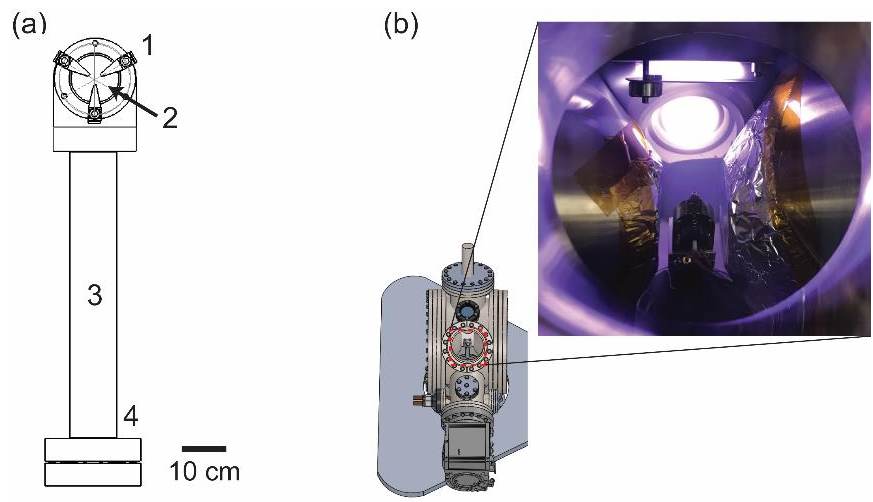

**Marks *et al.* Figure 4**

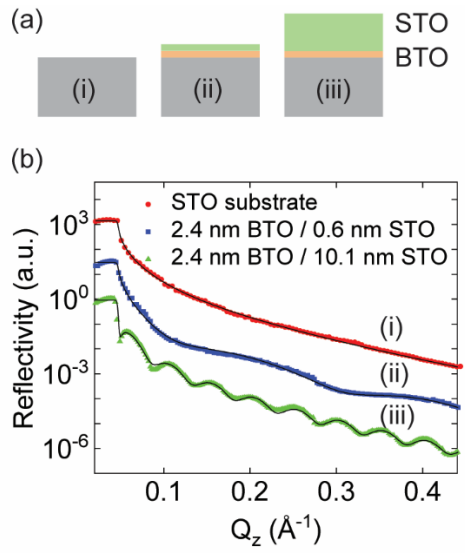

**Marks *et al.* Figure 5**

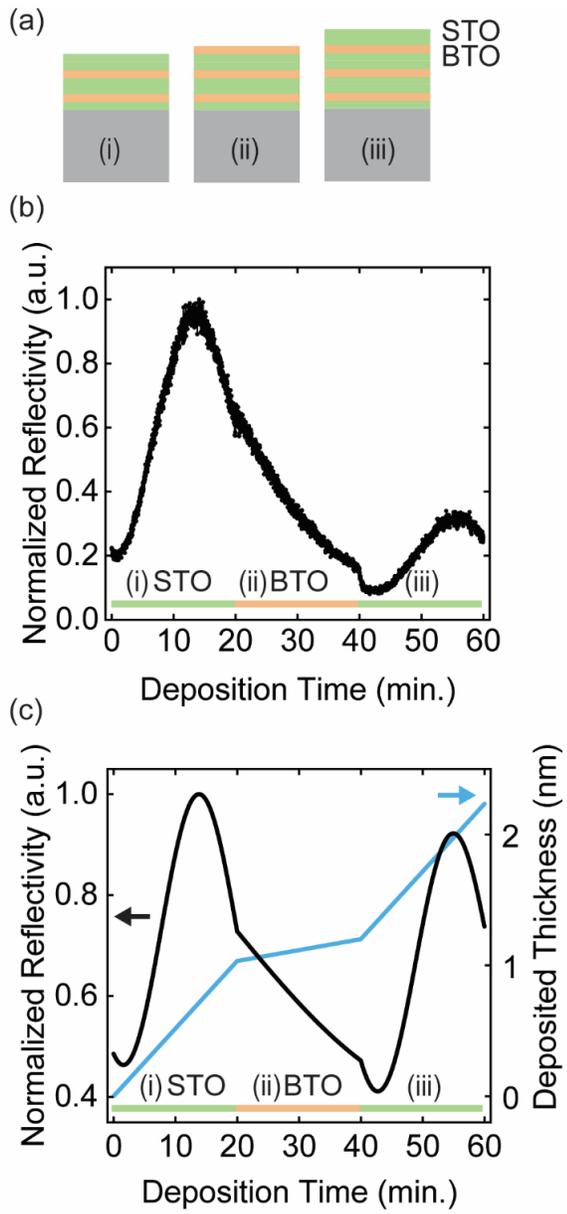

**Marks *et al.* Figure 6**

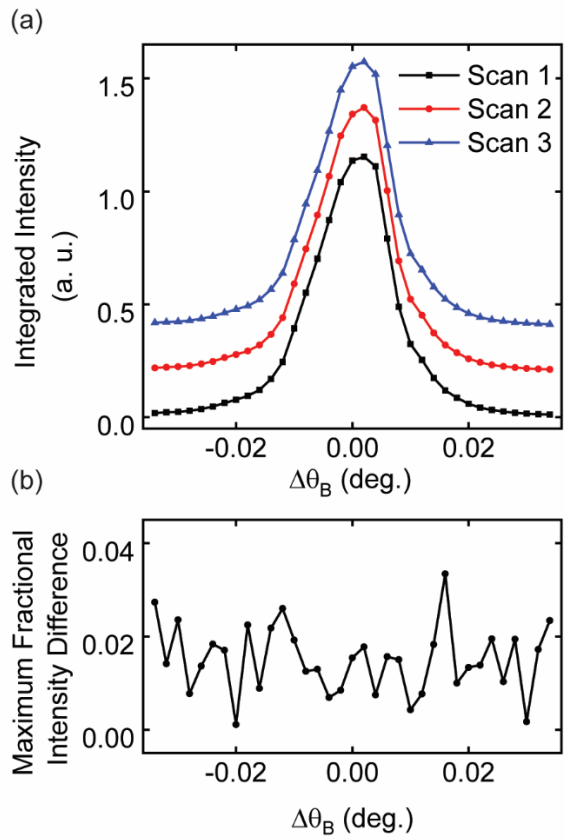